\documentclass[a5paper,pagesize,10pt,bibtotoc,pointlessnumbers,
normalheadings,DIV=9,twoside=false]{scrbook}

\KOMAoptions{DIV=last}

\usepackage{trajan}

\usepackage[utf8]{inputenc}
\usepackage[T1]{fontenc}

\usepackage[sc]{mathpazo}
\usepackage{verbatim} 
\usepackage{listings} 

\usepackage{blindtext}

\title{O \'{A}tomo}
\author{Francisco Caruso \& Vitor Oguri}
\date{\today}

\begin{document}

\begin{titlepage}
		\centering{
			{\fontsize{40}{48}\selectfont
			O \'{A}tomo}
		}\\
			
		\vspace{10mm}
		\centering{\Large{Francisco Caruso \& Vitor Oguri}}\\
		\vspace{\fill}
		\centering \large{PROFCEM Manuscript 2015}
\end{titlepage}

\newpage{}
\thispagestyle {empty}
\vspace*{2cm}

\begin{center}
	\Large{\parbox{10cm}{
		\begin{raggedright}
		{\Large
			\textit{Faz-se um apanhado hist\'{o}rico de algumas das principais contribui\c{c}\~{o}es de diferentes \'{a}reas da Ci\^{e}ncia com as quais, ao longo de s\'{e}culos, construiu-se uma vis\~{a}o cient\'{\i}fica, s\'{o}lida e consistente, do \'{a}tomo. Destacam-se os experimentos que nos levaram a uma s\'{e}rie de evid\^{e}ncias acerca da natureza composta, n\~{a}o elementar, do \'{a}tomo.}
		}
	
		\vspace{.5cm}\hfill{--- Francisco Caruso \& Vitor Oguri}
		\end{raggedright}
	}
}
\end{center}

\newpage

\section*{Pre\^{a}mbulo}\label{Int}
O atomismo filos\'{o}fico teve sua origem no per\'{\i}odo de florescimento da filosofia grega, mais precisamente com as contribui\-\c{c}\~{o}es de Leucipo e Dem\'{o}crito.  Segundo esse \'{u}ltimo, ``nada exis\-te al\'{e}m de \'{a}tomos e vazio; tudo mais \'{e} opini\~{a}o''. Devido \`{a} limita\c{c}\~{a}o de espa\c{c}o, n\~{a}o abordaremos aqui os aspectos mais especulativos acerca do \'{a}tomo, remetendo o leitor \`{a}s refer\^{e}ncias
\cite{FM}-\cite{Chalmers}. Nossa narrativa ter\'{a} in\'{\i}cio na contribui\c{c}\~{a}o de Newton ao desenvolvimento de uma ``nova'' Qu\'{\i}mica.

Ao reunificar a F\'{\i}sica, propondo que os movimentos celestes eram descritos pela mesma lei que regia a queda dos corpos na Terra -- a lei da Gravita\c{c}\~{a}o Universal -- Newton, de certa forma, atribui um car\'{a}ter especial \`{a} for\c{c}a ``peso''. N\~{a}o por acaso, alguns historiadores da ci\^{e}ncia defendem que a revolu\c{c}\~{a}o que o franc\^{e}s Antoine Laurent Lavoisier introduziu na Qu\'{\i}mica do s\'{e}culo XVIII tenha a ver com a f\'{e} que ele tinha na balan\c{c}a de precis\~{a}o. De fato, para o qu\'{\i}mico franc\^{e}s, toda mudan\c{c}a podia e devia ser explicada e mensurada. Se observarmos bem, o principal programa cient\'{\i}fico da Qu\'{\i}mica do s\'{e}culo XIX, a partir de John Dalton, foi medir sistematicamente os {\em pesos at\^{o}micos dos elementos qu\'{\i}micos}, o que acabou permitindo ao qu\'{\i}mico russo Dmitri Mendeleiev construir a sua famosa {\em Tabela Pe\-ri\-\'{o}dica} \cite{Jensen}. Todas as simetrias e regularidades nela contidas ti\-ve\-ram que aguardar d\'{e}cadas para poderem ser efetivamente com\-preendidas com base em uma vis\~{a}o atom\'{\i}stica da mat\'{e}ria~\cite{FM}.

\section*{A r\~{a} amb\'{\i}gua e a eletr\'{o}lise}
Por volta de 1780, o anatomista e m\'{e}dico italiano Luigi Galvani havia descoberto que quando se tocavam duas extremidades de um m\'{u}sculo de uma r\~{a} dissecada com metais diferentes este se contra\'{\i}a. Galvani atribuiu tal fen\^{o}meno a propriedades do pr\'{o}prio m\'{u}sculo, postulando a exist\^{e}ncia de uma {\em eletricidade ani\-mal} que, de alguma forma, se relacionaria com a {\em vida}. O f\'{\i}sico italiano Alessandro Volta  polemizou com Galvani durante d\'{e}cadas \cite{Pera}. Segundo Volta, o experimento com a r\~{a} nada tinha a ver com ela, mas, sim, com os dois metais diferentes. No final de 1799, para provar sua tese, Volta concluiu seu experimento com o que chamou, talvez n\~{a}o sem ironia, {\em \'{o}rg\~{a}o de ele\-tricidade artificial}, hoje conhecido como a {\em pilha voltaica}. Em 1800, os cientistas ingleses, William Nicholson e Anthony Carlisle constroem uma pilha e fazem a primeira eletr\'{o}lise da \'{a}gua. Este foi um marco experimental na compreens\~{a}o do \'{a}tomo. Com ele, mostra-se, pela primeira vez, que a eletricidade pode ser utilizada para decompor liga\c{c}\~{o}es qu\'{\i}micas. Ora, at\'{e} ent\~{a}o, pensava-se que as transforma\c{c}\~{o}es
qu\'{\i}micas eram devidas a ``for\c{c}as qu\'{\i}micas''. Com a eletr\'{o}lise viu-se que as for\c{c}as el\'{e}tricas s\~{a}o capazes de provocar rea\c{c}\~{o}es qu\'{\i}micas. Por associa\c{c}\~{a}o direta, pode-se imaginar que as for\c{c}as de liga\c{c}\~{o}es qu\'{\i}micas sejam de natureza el\'etrica. \'{E} o in\'{\i}cio da {\em eletroqu\'{\i}mica}.

O estudo quantitativo da eletr\'{o}lise foi empreendido pelo f\'{\i}si\-co e qu\'{\i}mico ingl\^{e}s Michael Faraday, que chegou \`{a}s conhecidas {\em leis de Faraday}. Essas leis, junto com a hip\'{o}tese at\^{o}mica, permitem antever uma estrutura at\^{o}mica para a eletricidade. De fato, d\'{e}cadas mais tarde, o irland\^{e}s George Johstone Stoney estimou o valor da carga elementar como algo da ordem de $10^{-20}$~C.  Esta carga seria, posteriormente, identificada como a carga do {\em el\'etron}, denomina\c{c}\~{a}o dada aos {\em \'{a}tomos de eletricidade} pelo pr\'{o}prio Stoney.
Essa interpreta\c{c}\~{a}o criou condi\c{c}\~{o}es para uma melhor compreens\~{a}o da natureza at\^{o}mica da eletricidade, principalmente devido a observa\c{c}\~{o}es de fen\^{o}menos resultantes de descargas el\'{e}tricas em gases rarefeitos.

 Discursando em homenagem a Faraday, o f\'{\i}sico e m\'{e}dico alem\~{a}o Hermann von Helmholtz destacou o que seria o resultado mais importante dos estudos sobre a eletr\'{o}lise com as seguintes palavras:
\begin{quotation}
\noindent
{\em Se aceitamos a hip\'{o}tese de que as subst\^{a}ncias elementares s\~{a}o compostas de \'{a}tomos, n\~{a}o podemos deixar de concluir que tamb\'{e}m  a eletricidade, tanto positiva quanto negativa, se
subdivide em por\c{c}\~{o}es elementares que se comportam como {\rm \'atomos de eletricidade.}}
\end{quotation}

James Clerk Maxwell referiu-se assim \`{a} relev\^{a}ncia dos estudos da eletr\'{o}lise:
\begin{quotation}
\noindent
{\em De todos os fen\^{o}menos el\'{e}tricos, a eletr\'{o}lise parece ser o que melhor nos oferece um maior discernimento sobre a verdadeira natureza da corrente el\'{e}trica, porque encontramos correntes de mat\'{e}ria ordin\'{a}ria e correntes de ele\-tri\-cidade formando partes essenciais do mesmo fen\^{o}meno.}
\end{quotation}

\section*{A espectroscopia}
Muitas das ideias sobre a estrutura at\^{o}mica e molecular que surgiram no in\'{\i}cio do s\'{e}culo~XX estavam, de certo modo, intimamente ligadas ao desenvolvimento da investiga\c{c}\~{a}o da radia\c{c}\~{a}o emitida pela mat\'{e}ria s\'{o}lida ou gasosa, gra\c{c}as ao trabalho pioneiro dos alem\~{a}es Robert Wilhelm Bunsen e Gustav Kirchhoff \cite{Kirchhoff-Bunsen}, a partir da inven\c{c}\~{a}o do espectr\'{o}grafo \'{o}ptico e do desenvolvimento, entre 1855 e 1863,  do que se convencionou chamar de  {\em espectroscopia}. Estudos nessa \'{a}rea permitiram a des\-coberta de novos elementos qu\'{\i}micos e levaram tamb\'{e}m o f\'{\i}sico estadunidense Albert Abraham Michelson a definir um novo padr\~{a}o para o metro: 1 m = 1\,553\,163,5 comprimentos de onda da linha vermelha do c\'{a}dmio. Deram tamb\'{e}m origem a uma nova \'{a}rea de investiga\c{c}\~{a}o astrof\'{\i}sica, que busca conhecer a composi\c{c}\~{a}o qu\'{\i}mica das estrelas.

Cada elemento qu\'{\i}mico d\'{a} origem a um espectro de emiss\~{a}o caracter\'{\i}stico, como se fosse uma esp\'{e}cie de ``impress\~{a}o digital'', \'{u}nica para cada elemento. Para os gases mono-at\^{o}micos, esses espectros, projetados em um anteparo ou visualizados por meio de um microsc\'{o}pio, apresentam-se, em geral, como um conjunto de {\em linhas} espa\c{c}adas e paralelas e, para os gases contendo dois ou mais \'{a}tomos, como {\em bandas} cont\'{\i}nuas. Sem d\'{u}vida, o espectro mais famoso acabou sendo o do \'{a}tomo de hidrog\^{e}nio, o elemento mais simples encontrado na Natureza. As regularidades desse espectro, descritas matematicamente pelo professor de Matem\'{a}tica e Latim, o su\'{\i}\c{c}o  Johann Jakob Balmer, foram de fundamental import\^{a}ncia para o sucesso do modelo de \'{a}tomo proposto pelo f\'{\i}sico dinamarqu\^{e}s Niels Bohr, j\'{a} no seculo~XX. Em \'{u}ltima an\'{a}lise, foi a chave para a compreens\~{a}o da natureza qu\^{a}ntica do \'{a}tomo \cite{FM}.

Observando o espectro de absor\c{c}\~{a}o na descarga el\'{e}trica entre eletrodos de carbono, iluminado com a luz do Sol, Foucault, em 1849, concluiu que a subst\^{a}ncia que emite luz de uma dada frequ\^{e}ncia tamb\'{e}m absorve melhor a luz nessa frequ\^{e}ncia. Essa conclus\~{a}o parece refor\c{c}ar a ideia de que os fen\^{o}menos de emiss\~{a}o e absor\c{c}\~{a}o seriam devidos a uma esp\'{e}cie de resson\^{a}ncia entre a radia\c{c}\~{a}o e os \'{a}tomos de uma subst\^{a}ncia, ou seja, sugere que os {\em \'{a}tomos} seriam {\em sistemas compostos}. Segundo Maxwell,
\begin{quotation}
\noindent
{\em foram essas observa\c{c}\~{o}es que primeiro levaram \`{a} conclus\~{a}o de que o espectro implicava que os \'{a}tomos tivessem estrutura, ou seja, fossem um sistema capaz de executar movimentos internos de vibra\c{c}\~{a}o.}
\end{quotation}

\section*{O experimento de Thomson}

Outras evid\^{e}ncias de uma subestrutura do \'{a}tomo foram obtidas a partir do surgimento dos chamados {\em tubos de Geissler}, {\em ampolas de Crookes}, ou ainda {\em tubos de raios
cat\'{o}dicos}, e dos estudos de novos fen\^{o}menos descobertos com esses tubos \cite{FM}. Vamos nos referia aqui apenas ao trabalho do f\'{\i}sico ingl\^{e}s Joseph John Thomson. Segundo ele,
\begin{quotation}
\noindent {\em Temos nos raios cat\'{o}dicos mat\'{e}ria em um novo estado, um estado no qual a subdivis\~{a}o da mat\'{e}ria \'{e} levada muito al\'{e}m do que no estado gasoso ordin\'{a}rio: um estado no qual toda mat\'{e}ria -- isto \'{e}, mat\'{e}ria derivada de diferentes fontes, como hidrog\^{e}nio, oxig\^{e}nio etc. -- \'{e} uma e do mesmo tipo; essa mat\'{e}ria \'{e} a subst\^{a}ncia da qual todos os elementos qu\'{\i}micos s\~{a}o feitos.}
\end{quotation}

Thomson havia estabelecido que os raios cat\'{o}dicos eram des\-viados tanto por campos magn\'{e}ticos como por campos ele\-trost\'{a}ticos. Usando esse fato e um tubo de raios cat\'{o}dicos ele foi capaz de calcular a raz\~{a}o $e/m$ entre a carga el\'{e}trica e a massa desses corp\'{u}sculos universais (os {\em el\'{e}trons}). Com essas medidas, constatou que a raz\~{a}o $e/m$ para os raios cat\'{o}dicos era aproximadamente 1\,840 vezes maior que a mesma raz\~{a}o para o hidrog\^{e}nio ionizado.

O estabelecimento do el\'{e}tron como constituinte subat\^{o}mico levou o pr\'{o}prio Thomson a propor um modelo f\'{\i}sico para o \'{a}tomo baseado em for\c{c}as eletrost\'{a}ticas. Estava, assim, definitivamente mostrado que o \'{a}tomo {\em n\~{a}o era indivis\'{\i}vel}, como proposto pelos gregos e at\'{e} ent\~{a}o aceito pelos qu\'{\i}micos. Restava, ainda, se chegar a um modelo at\^{o}mico coerente e est\'{a}vel. O caminho foi longo \cite{FM}-\cite{Pullman}.

\section*{O espectro de raios X e o n\'{u}mero de el\'{e}trons}

Uma outra descoberta que resultou do estudo emp\'{\i}rico envolvendo os tubos de raios cat\'{o}dicos foi a dos raios~X, pelo alem\~{a}o Wilhelm R\"{o}ntgen, quando os f\'{\i}sicos come\c{c}aram a
se perguntar se os raios cat\'{o}dicos se propagariam fora dos tubos. Em 1894, Philipp Lenard, ent\~{a}o assistente de Heinrich Hertz, idealizou um aparato, com o qual estudou o que aconteceria com os raios cat\'{o}dicos ao se propagarem no ar, fora do tubo. Com esse dispositivo, Lenard p\^{o}de observar que os raios cat\'{o}dicos se propagavam at\'{e} uma dist\^{a}ncia de poucos cent\'{\i}metros do tubo, n\~{a}o apenas no ar, mas tamb\'{e}m em outros gases. Verificou, ainda, que os raios eram capazes de impressionar chapas fotogr\'aficas e de tornar fluorescentes certos materiais, como, por exemplo, o platino-cianeto de b\'{a}rio, s\'{o}lido cristalino que apresenta tona\-li\-dades verde e amarela conforme a incid\^{e}ncia de luz que o ilumina. Foi utilizando um tubo de Lenard que R\"{o}ntgen se prop\^os a estudar, em novembro de 1895, a fluoresc\^{e}ncia de certas subs\-t\^{a}ncias. Para eliminar efeitos indesej\'aveis, R\"ontgen introduziu o tubo com o qual trabalharia em uma caixa de papel\~{a}o preto, de modo a bloquear raios vis\'{\i}veis e ultravioleta provenientes do tubo. Desse modo, apenas os raios cat\'{o}dicos passariam pela janela de Lenard, sendo colimados para a dire\c{c}\~{a}o dos objetos contendo as subst\^{a}ncias fluorescentes. Com a sala completamente escura, R\"{o}ntgen observou que um cart\~{a}o coberto por uma solu\c{c}\~{a}o de platino-cianeto de b\'{a}rio estava iluminado. Entretanto, os raios cat\'{o}dicos se propagam no ar por apenas alguns poucos cent\'{\i}metros, e o cart\~{a}o alvejado estava localizado a muito mais do que isso; cerca de 2~m.
Com o tubo isolado, qual seria a origem da fluoresc\^{e}ncia? Mais surpreendente ainda foi o fato de que  o papel n\~{a}o estava na linha do feixe de raios cat\'{o}dicos. O que provocava, ent\~{a}o, aquela luminesc\^{e}ncia? Intrigado e perplexo com sua origem desconhecida, R\"{o}ntgen deu a esses raios o nome provis\'orio de raios~X -- baseado na letra normalmente atribu\'{\i}da \`{a} inc\'{o}gnita de um problema a resolver -- nome este que passou a ser definitivamente adotado.

Poucas descobertas de f\'{\i}sica b\'{a}sica tiveram aplica\c{c}\~{o}es pr\'{a}ticas t\~{a}o cedo e espetaculares. Os raios X permitiram aos m\'{e}dicos ``verem'' dentro dos corpos sem ter que abri-los. Deu tamb\'{e}m enorme impulso \`{a} \'{a}rea da {\em Cristalografia} e permitiu  a confirma\c{c}\~{a}o de que a mat\'{e}ria em estado cristalino seria um arranjo regular de \'{a}tomos e mol\'{e}culas dispostos em camadas, respaldando, portanto, a vis\~{a}o atom\'{\i}stica da mat\'{e}ria. Ainda do ponto de vista da F\'{\i}sica b\'{a}sica, em 1904, Charles Barkla dedicou-se a determinar o n\'{u}mero de el\'{e}trons contidos em \'{a}tomos-alvo em processos de espalhamento de raios~X. Somente em 1911, com dados mais precisos, Barkla foi capaz de mostrar que, para \'{a}tomos leves, o n\'{u}mero de el\'{e}trons \'{e} a metade do n\'{u}mero de massa do correspondente elemento qu\'{\i}mico. Por fim, no bi\^{e}nio 1913-1914, os trabalhos de Henry Moseley com espectros de raios~X confirmam algumas das ideias de Rutherford e de Bohr sobre a constitui\c{c}\~{a}o at\^{o}mica da mat\'{e}ria \cite{FM}.

Os raios~X foram tamb\'{e}m relevantes no estudo do chamado {\em efeito Compton}, o qual ser\'{a} discutido adiante no texto.

\section*{A radioatividade}
O f\'{\i}sico franc\^{e}s Henry Becquerel, ap\'{o}s tomar conhecimento dos trabalhos de R\"{o}ntgen, passou a investigar se algumas subst\^{a}ncias, que se tornavam fosforescentes sob a incid\^{e}ncia de luz, eram capazes de emitir qualquer tipo de radia\c{c}\~{a}o penetrante, com os raios~X. Descobriu, assim, os {\em raios ur\^{a}nicos}. O que mais o intrigava era a natureza {\em espont\^{a}nea} da emiss\~{a}o desses raios, aparentemente sem causas externas. Essa quest\~{a}o mobilizou muitos f\'{\i}sicos e s\'{o} foi compreendida d\'{e}cadas mais tarde, com a Mec\^{a}nica Qu\^{a}ntica.

O casal Curie dedicou-se muito ao estudo dessas emiss\~{o}es, e Marie Curie cunhou o termo {\em Radioatividade}.  De forma muito resumida, podemos dizer desses estudos, com a descoberta dos raios $\alpha$, $\beta$ e $\gamma$, que a radioatividade natural possu\'{\i}a duas componentes formadas de part\'{\i}culas carregadas e uma terceira ($\gamma$) de natureza eletromagn\'{e}tica.

Segundo Rutherford, a ``grande semelhan\c{c}a das mudan\c{c}as no r\'{a}dio, t\'{o}rio e act\'{\i}nio \'{e} muito not\'{a}vel e indica alguma peculiaridade da constitui\c{c}\~{a}o at\^{o}mica que ainda est\'{a} por ser elucidada''. Al\'{e}m disso, havia uma pergunta relevante sem resposta: por que todos os \'{a}tomos de um certo elemento n\~{a}o decaem no mesmo tempo j\'{a} que todos eles s\~{a}o id\^{e}nticos entre si?

A resposta a essa quest\~{a}o s\'{o} viria a partir de um novo olhar sobre o microcosmo que ainda estava por vir com a Mec\^{a}nica Qu\^{a}ntica. Mas mesmo sem a resposta, a {\em lei dos decaimentos radioativos} permitiram que Ernest Rutherford calculasse o {\em n\'{u}mero de Avogadro} de um modo at\'{e} ent\~{a}o inesperado, confirmando a natureza molecular da mat\'{e}ria. Nesse sentido, cabe tamb\'{e}m recordar o sucesso da interpreta\c{c}\~{a}o estat\'{\i}stica da Teoria Cin\'{e}tica dos Gases \cite{Brush}.

\section*{O movimento browniano}
O movimento browniano tornou-se, no in\'{\i}cio do s\'{e}culo~XX, uma das mais convincentes provas acerca da realidade das mo\-l\'{e}\-cu\-las, ou seja, da hip\'{o}tese corpuscular da mat\'{e}ria. Isto porque o f\'{\i}sico franc\^{e}s Jean Perrin, em uma s\'{e}rie de trabalhos sistem\'{a}ticos, mediu o n\'{u}mero de Avogadro de muitas formas di\-fe\-ren\-tes, encontrando sempre resultados compat\'{\i}veis. Se ainda havia qualquer d\'{u}vida acerca da natureza molecular da mat\'{e}ria (como para o qu\'{\i}mico alem\~{a}o Friedrich Wilhelm Ostwald) ela se dissipou com os resultados obtidos por Perrin.

Provavelmente nenhuma outra constante fundamental despertou o interesse de tantos f\'{\i}sicos do porte de Amp\`{e}re, Losch\-midt, Max\-well, Boltzmann, Thomson, Rayleigh, Planck, Einstein, Rutherford, Millikan, Perrin e outros, quanto o {\em n\'{u}mero de Avogadro}. Esse fato por si s\'{o} j\'{a} sugere a for\c{c}a da concep\c{c}\~{a}o at\^{o}mica da mat\'{e}ria e seu papel basilar na cons\-tru\c{c}\~{a}o do co\-nhe\-cimento cient\'{\i}fico moderno que, concluindo, podem muito bem ser resumidos nas palavras do f\'{\i}sico americano Richard Feynman:
 \begin{quotation}
\noindent
{\em Se, em algum cataclismo, todo o conhecimento cient\'{\i}fico fosse destru\'{\i}do e somente uma senten\c{c}a fosse transmitida para as pr\'{o}ximas ge\-ra\-\c{c}\~{o}es de cria\-tu\-ras, que e\-nun\-cia\-do con\-te\-ria mais in\-for\-ma\c{c}\~{a}o em me\-nos pa\-la\-vras? Acre\-di\-to que se\-ja a {\rm hi\-p\'o\-te\-se at\^o\-mi\-ca} (...) de que {\em to\-das as coi\-sas s\~ao fei\-tas de \'{a}tomos} (...). Nessa \'{u}nica sen\-ten\-\c{c}a, vo\-c\^{e} ver\'{a}, exis\-te uma {\rm e\-nor\-me} quan\-ti\-da\-de de in\-for\-ma\-\c{c}\~{o}es sobre o mundo.}
\end{quotation}

\section*{O experimento de Rutherford}

O experimento de espalhamento de part\'{\i}culas $\alpha$ por alvos delgados serviu para que Rutherford introduzisse o importante conceito de {\em n\'{u}cleo at\^{o}mico}. Sua conclus\~{a}o foi de que o \'{a}tomo tem uma enorme regi\~{a}o vazia. H\'{a} um n\'{u}cleo ocupando uma pequena regi\~{a}o da ordem de $10^{-12}$~cm, enquanto o el\'{e}tron na \'{o}rbita mais pr\'{o}xima descreve uma trajet\'{o}ria circular com raio da ordem de $10^{-8}$~cm. Em outras palavras, em uma escala na qual o n\'{u}cleo at\^{o}mico tivesse um raio de cerca de 1~m, o el\'{e}tron mais pr\'{o}ximo estaria a 10~km de dist\^{a}ncia. Essa nova vis\~{a}o abrir\'{a} caminho para uma nova \'{a}rea de pesquisa em F\'{\i}sica B\'{a}sica: a {\em F\'{\i}sica Nuclear}. A hist\'{o}ria das tentativas de conhecer e dominar o n\'{u}cleo at\^{o}mico todos conhecem. Por um lado, h\'{a} a triste inven\c{c}\~{a}o da bomba at\^{o}mica, e de outras armas antes impens\'{a}veis, mas tamb\'{e}m trouxe, por outro lado, a medicina nuclear, com grande impacto social.

\section*{O efeito Compton}

O {\em efeito Compton} foi, na realidade, a evid\^{e}ncia experimental que faltava para que a comunidade cient\'{\i}fica admitisse a exist\^{e}ncia do {\em f\'{o}ton} como constituinte da luz, colocando um ponto final nessa quest\~{a}o que foi assunto de grande pol\^{e}mica.  O que era aceito at\'{e} ent\~{a}o pela maioria dos f\'{\i}sicos era o resultado das discuss\~{o}es surgidas na Confer\^{e}ncia de Solvay, de 1911, onde foi aceita apenas a descontinuidade na emiss\~{a}o e na absor\c{c}\~{a}o de luz, e n\~{a}o da pr\'{o}pria energia da luz, como havia proposto Einstein.

De fato, Compton mostrou que o espalhamento de raios~X pela mat\'{e}ria resulta da colis\~{a}o de f\'{o}tons com el\'{e}trons praticamente livres no interior da mat\'{e}ria. Sua compreens\~{a}o se constituiu em um argumento definitivo em favor da ideia de quantiza\c{c}\~{a}o da radia\c{c}\~{a}o, ou seja, da exist\^encia de \textit{f\'{o}tons}.



\section*{Coment\'{a}rios finais}\label{conclusions}

Em 1932, chegou-se a um quadro conceitual para explicar a mat\'{e}ria e a luz, constitu\'{\i}do de quatro {\em part\'{\i}culas elementares}: o el\'{e}tron, o pr\'{o}ton e o n\^{e}utron (os constituintes do \'{a}tomo) e o f\'{o}ton, ({\em quantum} da luz). Portanto, a F\'{\i}sica nos levou a um \'{a}tomo novo, dotado de uma estrutura f\'{\i}sica e n\~{a}o mais indivis\'{\i}vel e imut\'{a}vel. Sua descri\c{c}\~{a}o, do ponto de vista te\'{o}rico, exigiu o desenvolvimento de uma nova teoria: a {\em Mec\^{a}nica Qu\^{a}ntica}, que inclui conceitos novos como o de {\em spin} \cite{Tomonaga}. Sem esse conceito e sem esse novo referencial te\'{o}rico, n\~{a}o se pode compreender, finalmente, as regularidades da Tabela Peri\'{o}dica em termos da distribui\c{c}\~{a}o eletr\^{o}nica dos \'{a}tomos.

Sugerimos que o leitor interessado em ler mais sobre essa hist\'{o}ria consulte nosso livro citado na refer\^{e}ncia \cite{FM}, no qual as bases dessa nova mec\^{a}nica s\~{a}o apresentadas em uma pers\-pectiva hist\'{o}rica. O que procuramos fazer aqui, conciliando a sugest\~{a}o de tema feita pelos organizadores do evento com a limita\c{c}\~{a}o de espa\c{c}o, foi apenas mostrar o qu\~{a}o rica e intricada foi a hist\'{o}ria da compreens\~{a}o do {\em \'{a}tomo} e que, em grande medida, v\'{a}rios resultados experimentais foram cruciais nesse processo de busca, para o qual n\~{a}o se vislumbra um horizonte.

\end{document}